\documentstyle[a4,11pt,epsf]{article}
\epsfverbosetrue
\newcommand{\beq}{\begin{equation}}
\newcommand{\eeq}{\end{equation}}   

 0

\begin{document}

\begin{titlepage}
 
%This is for the preprint number
\begin{flushright}
Dublin Preprint: TCDMATH 98-14\\
%cond-mat/9812004\\
%hep-lat/9812004\\

\end{flushright}

\vspace{5mm}
 
\begin{center}
{\huge The Phase Structure of the Weakly
  \\[3mm]
Coupled Lattice Schwinger Model}\\[15mm]
{\bf R. Kenna\footnote{Supported by EU TMR
Project No. ERBFMBI-CT96-1757 and a Forbairt European Presidency 
Post Doctoral Fellowship.},
\\
School of Mathematics, Trinity College Dublin, Ireland\\
~\\ 
C. Pinto and J.C. Sexton
\\
School of Mathematics, Trinity College Dublin, Ireland\\
and \\
Hitachi Dublin Laboratory, Dublin, Ireland
} 
\\[3mm]~\\ 
December 1998
\end{center}
\begin{abstract}
The weak coupling expansion is applied to the single flavour
Schwinger model with Wilson fermions
on a symmetric toroidal lattice of finite extent.
We develop a new analytic
method which permits the expression of the partition 
function as a product of pure gauge expectation values whose 
zeroes are the Lee--Yang zeroes of the model. Application of
standard finite--size scaling techniques to these zeroes
recovers previous numerical results for the small and moderate lattice
sizes to which those studies were restricted. Our techniques,
employable for arbitrarily large lattices, reveal the absence
of accumulation of these zeroes on the real hopping parameter axis
at constant weak gauge coupling.
The consequence of this previously unobserved behaviour is the
absence of a zero fermion mass
phase transition in the Schwinger model with 
single flavour Wilson fermions at constant weak gauge coupling.
\end{abstract}
\end{titlepage}

\newpage

%%%%%%%%%%%%%%%%%%%%%%%%%%%%%%%%%%%%%%%%%%%%%%%%%%%%%%%%%%%%%%%%%%%%%%%%
%%%%%%%%%%%%%%%%%%%%%%%%%%%%%%%%%%%%%%%%%%%%%%%%%%%%%%%%%%%%%%%%%%%%%%%%
\paragraph{Introduction}
\setcounter{equation}{0}
%%%%%%%%%%%%%%%%%%%%%%%%%%%%%%%%%%%%%%%%%%%%%%%%%%%%%%%%%%%%%%%%%%%%%%%%
%%%%%%%%%%%%%%%%%%%%%%%%%%%%%%%%%%%%%%%%%%%%%%%%%%%%%%%%%%%%%%%%%%%%%%%%

In lattice gauge theory, there has been considerable discussion
on the phase structure of gauge theories with Wilson fermions.
A system of free Wilson fermions exhibits a second order phase transition
at $(\beta,\kappa)=(\infty,1/(2d))$, $\beta$ and $\kappa$ being
the usual inverse gauge coupling squared and hopping parameters 
respectively
and $d$ being the lattice dimensionality.
Recent discussions concern the extent to which this phase
transition extends into the $(\beta,\kappa)$ plane, the
expectation being that there is a line of phase transitions (the
chiral limit) extending to the strong coupling limit $\beta = 0$.
In the simplest model to which this applies, the lattice Schwinger 
model \cite{Sc62}, there is also 
a second order phase transition in the strongly
coupled limit \cite{GaLa95Karsch}.
This critical line, which is expected to connect these two phase 
transitions,
is also expected to recover massless physics
and therein lies the importance in determining its nature and position.

While earlier numerical results casted doubt on the above
scenario \cite{GaLa92},
more recent ones support the hypothesis that the critical line 
extends
over the whole range of couplings \cite{GaLa94,HiLa98,AzDi96}.
To our knowledge, only two groups have attempted to determine the
phase diagram in the lattice Schwinger model. While in rough
agreement regarding the location of the critical line,
they differ in their conclusions regarding its quantitive critical
properties.
Using Lee--Yang zeroes
\cite{LY}, finite--size scaling  techniques \cite{IPZ} as well as PCAC 
relations,
the results of \cite{GaLa94,HiLa98} 
support the free boson scenario, where the
model lies in the same universality class of the Ising model,
with correlation length critical exponent $\nu = 1$ and the specific
heat (chiral susceptibility) exponent $\alpha = 0$. This is not in
agreement with \cite{AzDi96} where finite--size scaling
 of Lee--Yang zeroes on larger
lattices provides evidence for $\nu = 2/3$, $\alpha = 2/3$.

The precise analytical determination of  the 
phase structure in the weakly coupled regime is 
the primary motivation for this work. We contend that the Lee--Yang 
zeroes
in the Schwinger model with fixed weak gauge coupling
display very unusual size--dependent
behaviour, do not accumulate on the real hopping parameter axis and
may not, in fact, lead to a phase transition. 
The behaviour of these zeroes 
is, however, such as to mimic the appearance
of a phase transition when a finite--size analysis is restricted to
small or moderate lattice sizes.

%%%%%%%%%%%%%%%%%%%%%%%%%%%%%%%%%%%%%%%%%%%%%%%%%%%%%%%%%%%%%%%%%%%%%%%%
%%%%%%%%%%%%%%%%%%%%%%%%%%%%%%%%%%%%%%%%%%%%%%%%%%%%%%%%%%%%%%%%%%%%%%%%
\paragraph{Lattice Schwinger Model}
\setcounter{equation}{0}
%%%%%%%%%%%%%%%%%%%%%%%%%%%%%%%%%%%%%%%%%%%%%%%%%%%%%%%%%%%%%%%%%%%%%%%%
%%%%%%%%%%%%%%%%%%%%%%%%%%%%%%%%%%%%%%%%%%%%%%%%%%%%%%%%%%%%%%%%%%%%%%%%

We consider a finite $d=2$ dimensional lattice with spacing $a$
and $N$ sites in each direction. For the fermion fields, 
we impose antiperiodic boundary 
conditions in the temporal ($1$-) direction and periodic boundary 
conditions in the spatial ($2$-) direction. Lattice sites are labelled 
$x_\mu = n_\mu a$ where $ n_\mu = -N/2, \dots, N/2 - 1$. 
With these mixed boundary conditions the momenta for the
Fourier transformed fermion fields  are
$ k_\mu = 2\pi {\hat{k}}_\mu /Na $,
where $ {\hat{k}}_1 = -N/2+1/2,  \dots, N/2 - 1/2$ and
$ {\hat{k}}_2  =  -N/2,  \dots, N/2 - 1$.

The action for gauge invariant lattice QED with Wilson fermions
is $ S_{\rm{QED}} = S_G[\phi] + S_F^{(W)}[\phi,\psi,\bar{\psi}]$,
where $
 S_F^{(W)} [\phi,\psi,\bar{\psi}] = \sum_{m,n} 
\bar{\psi} (m) M^{(W)}(m,n)  \psi (n)
 $ and
\begin{equation}
 M^{(W)}(m,n) =
 \frac{1}{2\kappa}          \delta_{m,n} 
 -
 \frac{1}{2}\sum_\mu{
 \left\{
         ( r - \gamma_\mu ) U_\mu(m)\delta_{m+\hat{\mu},n}
       + ( r + \gamma_\mu ) U^\dagger_\mu(n)\delta_{m-\hat{\mu},n}
 \right\}
}
\quad .
\label{3.5}
\end{equation}
This Wilson fermion matrix
can be written in terms of free and interacting parts
as $M^{(W)} = M^{(0)} + M^{({\rm{int}})} $ where
$M^{(0)}$ is given by $U_\mu = 1$ in (\ref{3.5}).
Here
$U_\mu (n) = \exp{i \phi_\mu(n)} = \exp{(i e_0 a A_\mu (n))}$
is the link variable. The Wilson parameter, $r$, is henceforth
set to unity.
The partition function is $
 Z_{\rm{QED}} = 
 \int{
      {\cal{D}} U {\cal{D}} \bar{\psi} {\cal{D}} \psi
      \exp{(
         \{
           -S_{\rm{G}} -S^{(W)}_{\rm{F}}
         \}
        )}
     }
\propto
 \int{
      {\cal{D}} U
     \exp{
          \{-S_{\rm{G}}
          \} 
         } 
     \det{M^{(W)}}
      }
\propto
 \langle
       \det{M^{(W)}}
 \rangle
$,
the integration over the Grassmann Fermionic fields having been 
performed.
Here $\langle {\cal{O}} \rangle $ is the pure gauge expectation 
value of the quantity ${\cal{O}}$ and the pure gauge action is
$
 S_G[\phi]=
    \beta\sum_P{\left[1-\frac{1}{2}(U_P+U_P^\dagger)\right]}
$,
where $U_P$ is the usual product of link variables around a 
fundamental plaquette and where $\beta = 1/e_0^2 a^2$.
In the weak coupling expansion we
employ the Feynman gauge for the calculation of pure gauge
expectation values. (There, for large enough $N$, the gauge 
propagator
in momentum space is given approximately by
$\langle \tilde{\phi}_\mu (\hat{p}) \tilde{\phi}_\nu (-\hat{p})\rangle
= \delta_{\mu \nu} (N^d a^{2d}/2 \beta)
 \sum_{\rho,p\ne 0}{(1 -\cos{p_\rho a})}$
with the infra--red zero--momentum
mode excluded as standard \cite{irzm}.)

For $\beta = \infty$, the only gauge configurations which contribute
have $S_G = 0$.
We assume for the following analysis that all gauge
configurations satisfying this condition belong to a single equivalence
class under the group of gauge transformations. Since the fermion
determinant is gauge invariant under the same group of gauge 
transformations
it is sufficient to evaluate the full partition function on this single
configuration to determine the behaviour of the system at 
$\beta = \infty$
\cite{GaLa94}.
Therefore the partition function is
simply proportional to $ \det{M^{(0)}} $. 
Employing lattice Fourier transforms, the free fermion partition 
function can
be written in terms of its $dN^d$  eigenvalues by solving 
the $d\times d$ dimensional eigenvalue problem of the form
$M^{(0)}(p) | \lambda^{(0)}_\alpha (p) \rangle = \lambda^{(0)}_\alpha (p)
 | \lambda^{(0)}_\alpha (p) \rangle $ where $\alpha=1,\dots,d$.
In $d=2$ dimensions, the resulting eigenvalues are
$
 \lambda^{(0)}_\alpha (p) 
 = 
 1/2\kappa
 - \sum_{\rho=1}^2  \cos{p_\mu a} 
 + i (-)^\alpha  \sqrt{ \sum_{\rho=1}^2{\sin^2{p_\rho a}}}
$.

With mixed  boundary conditions,
the eigenvalues in the free fermion
case are either two--fold ($\hat{p}_2 = 0$ or $-N/2$) or four--fold 
degenerate.
In this case, and with $\eta = 1/2\kappa$, the Lee--Yang zeroes are 
 $\eta^{(0)}_\alpha(p) =
 \sum_{\rho=1}^2 r \cos{p_\mu a} 
 - i (-)^\alpha  \sqrt{\sum_{\rho=1}^2{\sin{p_\rho a}}}
$.
The lowest zero with finite real part in $\kappa$ corresponds
to $\hat{p} = (\pm 1/2,0)$ and for a symmetric lattice is
$ \eta^{(0)}_\alpha (\pm 1/2,0)
  = 
 1 +  \cos( \pi/N) 
 - i (-)^\alpha  {\sin{( \pi/N)}}$.
Pinching of the positive real finite
hopping parameter axis occurs at the massless point
$\kappa_c = 1/2d $ and application of finite--size scaling
 \cite{IPZ} to
the imaginary parts of the zero 
gives the critical exponent $ \nu = 1$ ($\alpha = 0$ then follows from
hyperscaling). Therefore the free fermion model is in the same
universality class as the Ising model in two dimensions and
describes free bosons.
In the presence of a gauge field the position and nature
of this phase transition will change and a quantitive
investigation into the nature of that change is the central aim of
this paper.

%%%%%%%%%%%%%%%%%%%%%%%%%%%%%%%%%%%%%%%%%%%%%%%%%%%%%%%%%%%%%%%%%%%%%%%%
%%%%%%%%%%%%%%%%%%%%%%%%%%%%%%%%%%%%%%%%%%%%%%%%%%%%%%%%%%%%%%%%%%%%%%%%
\paragraph{The Weak Coupling Expansion}
%\setcounter{equation}{0}
%%%%%%%%%%%%%%%%%%%%%%%%%%%%%%%%%%%%%%%%%%%%%%%%%%%%%%%%%%%%%%%%%%%%%%%%
%%%%%%%%%%%%%%%%%%%%%%%%%%%%%%%%%%%%%%%%%%%%%%%%%%%%%%%%%%%%%%%%%%%%%%%%

The weak coupling expansion is formally obtained by expanding the link
variables $U_\mu (n)$ as a power series in $e_0$.
The weak coupling expansion of the matrix 
$M^{({\rm{int}})}$ in (\ref{3.5})
is $ M^{({\rm{int}})} = M^{(1)} + M^{(2)} + 
{\cal{O}}(e_0^3)$ where $M^{(n)}$ is of order $e_0^n$.
The fermion determinant 
$\det{M^{(W)}} = \det{M^{(0)}} \times \det{{M^{(0)}}^{-1} M^{(W)}}
= \det{M^{(0)}} \exp{ {\rm{tr}} 
\ln{(1 + {M^{(0)}}^{-1} M^{({\rm{int}})})}}$
can then be expanded giving the following
`traditional' additive form for the weak coupling expansion 
\cite{Rothe}:
\begin{equation}
 \frac{ 
       \langle \det{M^{(W)}} \rangle 
      }{
       \det{M^{(0)}}
      }
 =
 1
 +
 \sum_{i=1}^{dN^d}{
                   \frac{t_i}{\eta-\eta_i^{(0)}}
                  } 
 -
 \frac{1}{2}
 \sum_{i\ne j}^{dN^d}{
                      \frac{
                            t_{ij}
                          }{
                            (\eta-\eta_i^{(0)})
                            (\eta-\eta_j^{(0)})
                           } 
                     }
 + \dots
\quad .
\label{additive}
\end{equation}
in which
$ t_i =  \langle M_{ii}^{({\rm{int}})} \rangle $ and
$ t_{ij} = t_{ji} 
= \langle M_{ij}^{({\rm{int}})} M_{ji}^{({\rm{int}})}\rangle 
-
\langle M_{ii}^{({\rm{int}})} M_{jj}^{({\rm{int}})}\rangle 
$. Here the index $i$ generically stands
for the combination of Dirac
index and momenta $(\alpha,p)$ which
label fermionic matrix elements so that $M_{ij}^{({\rm{int}})}$
represents $\langle \lambda_\alpha^{(0)} (q) | M^{({\rm{int}})}(q,p)
| \lambda_\beta^{(0)} (p) \rangle $.
We note at this point that the expansion is analytic in $\eta$ with
poles at $\eta = \eta_i^{(0)}$. 
We note further that $t_i=\langle M_{ii}^{({\rm{int}})} \rangle$ 
at order $e$ is proportional to the pure gauge expectation value of
the gauge field $\phi_\mu$ and is zero.

An alternative formulation of the partition function may be obtained
by writing the Wilson fermion matrix as $M^{(W)} = \eta 
+ H $ where $H$ is the hopping matrix. The fermion
determinant $\det M^{(W)} = \det(\eta + H)$, is a polynomial
in $\eta$ since for finite lattice size these matrices are
of finite dimension. Therefore the pure gauge expectation value
of the fermion determinant is also a polynomial in $\eta$ and as such
may be written
$
 \langle \det M^{(W)} \rangle = \prod_{i=1}^{dN^d}
 (\eta - \eta_i)
$. Here $\eta_i$ represent $\eta_\alpha (p)$ and are the Lee--Yang 
zeroes
in the complex $1/2 \kappa$ plane.
We may write a `multiplicative' weak coupling expansion as
\begin{equation}
 \frac{ 
     \langle  \det{M^{(W)}} \rangle 
      }{
       \det{M^{(0)}}
      }
 = \prod_{i=1}^{dN^d}
 \frac{(\eta - \eta_i)}{(\eta - \eta_i^{(0)})}
 = \prod_{i=1}^{dN^d}\left(
  1 - \frac{\Delta_i}{\eta - \eta_i^{(0)}}
                    \right)
\quad ,
\label{multiplicative}
\end{equation}
where $\Delta_i = \eta_i - \eta_i^{(0)}$ are the shifts that occur
in the zeroes when the gauge field is turned on and 
are the quantities to be determined.
Note again that (\ref{multiplicative}) is analytic in $\eta$ with
poles at $\eta_i^{(0)}$.
Since  the two expressions (\ref{additive}) and
(\ref{multiplicative}) 
must be equal, the residues of the poles
must be identical.

Let $t_i = t_i^{(2)} + \dots$,
$t_{ij} = t_{ij}^{(2)} + \dots$ 
and $\Delta_i = \eta_i^{(1)} + \eta_i^{(2)}
+ \dots $ where $\eta_i^{(1)}$ is ${\cal{O}}(e)$
and $t_i^{(2)}$, $t_{ij}^{(2)} $ and 
$\eta_i^{(2)}$ are ${\cal{O}}(e^2)$.
Assume the $n_j^{\rm{th}}$ free eigenvalues are degenerate for 
$j=1,\dots,D_n$ where $D_n$ is $2$ or $4$
and $\lambda^{(0)}_{n_j} \equiv 
\lambda^{(0)}_{n}$. At this stage we make no assumptions regarding 
the other eigenvalues. The traditional weak coupling expansion
 (\ref{additive}) and its
`multiplicative' counterpart (\ref{multiplicative}) have
single poles at $\eta = \eta^{(0)}_n$, the residues of which 
are easily calculated to $ {\cal{O}}( e^2 )$.
Equating the two residues order by order in $e$ gives
\begin{equation}
 \sum_{j=1}^{D_n}{ \eta_{n_j}^{(1)} }  =  0
\quad , \quad \quad \quad
 \sum_{j=1}^{D_n}{ \eta_{n_j}^{(2)} }
 =  
 D_n \bar{\Delta}_n^{(2)}
 \equiv
 \sum_{j=1}^{D_n}{ \left[
                     -  t_{n_j}^{(2)}
                     +
                     \sum_{i, \lambda_i^{(0)}\neq \lambda_n^{(0)} }{
                              \frac{t_{i n_j}^{(2)} }{
                              \eta_n^{(0)}-\eta_i^{(0)}
                               }
                            }
               \right]
            }
 \quad .
\label{spe2}
\end{equation}
In this latter equation, $ \bar{\Delta}_n^{(2)}$ is
the average of two zeroes which in 
the free fermion case were two--fold degenerate and is ${\cal{O}}(e^2)$.
Equating the residues of the corresponding double poles gives no extra
information at this order.

In the non--degenerate case $D_n=1$, (\ref{spe2}) is sufficient 
and necessary
to fully determine the shifts in the positions of the zeroes 
$\eta_{n_j}^{(1)}$ and $\eta_{n_j}^{(2)}$. This then amounts to a
new method to uniquely express the partition function and all derivable 
thermodynamic functions in multiplicative (as opposed to the
traditional `additive') form. Thus, we have expressed the pure gauge
expectation value of a product of fermion eigenvalues as
a product of pure gauge expectation values.

In the $D_n$--fold degenerate case, (\ref{spe2}) gives for 
the expansion  of the `multiplicative'
weak coupling expression (\ref{multiplicative}) to ${\cal{O}}(e^2)$,
\begin{equation}
 \prod_i{\left(
                1 - \frac{\Delta_i}{\eta - \eta_{n_i}^{(0)}}
        \right)}
 =
 1 + \sum_i{\frac{t_i^{(2)}}{\lambda_i^{(0)}}}
 +
 \sum_{i,j,\lambda_i^{(0)}\neq \lambda_j^{(0)}}{
 \frac{t_{ij}^{(2)}}{\lambda_i^{(0)}(\lambda_i^{(0)}-\lambda_j^{(0)})}
 +\sum_{i \neq j}{
 \frac{\Delta_i \Delta_j}{\lambda_i^{(0)} \lambda_j^{(0)}}
 }
 + \dots
 }
\quad .
\end{equation}
This is identical to the traditional weak coupling expansion
(\ref{additive}) to ${\cal{O}}(e^2)$ if
\begin{equation}
 \sum_i{\left( \frac{\eta_i^{(1)}}{\lambda_i^{(0)}} \right)^2}
 =
 \sum_{
       i\neq j,\lambda_i^{(0)}=\lambda_j^{(0)}
      }{
       \frac{t_{ij}^{(2)}}{\left(\lambda_i^{(0)}\right)^2 }
       }
\quad .
\label{MVAe2'}
\end{equation}
Setting $\eta=\eta_n^{(0)} + \epsilon$ where $\epsilon \neq 0$
and separating out the contributions coming from the $D_n$--fold 
degenerate
cases $n_j$ now gives that
\begin{equation}
 \sum_{i=1}^{D_n} (\eta_{n_i}^{(1)})^2 
 = 
 \sum_{i=1}^{D_n} \sum_{j=1,j\neq i}^{D_n}
 t^{(2)}_{n_i n_j} + {\cal{O}}(\epsilon)
\quad .
\label{eta1-4}
\end{equation}
Now the `multiplicative' and `additive' expressions for the partition
function coincide up to ${\cal{O}}(e^2)$ everywhere in the complex
$1/2 \kappa $ plane arbitrarily close to any pole and have the same
poles and the same residues (up to ${\cal{O}}(e^2)$) at those poles.
The partition function zeroes are
$
 \eta_{n_i} = \eta_n^{(0)} + \eta_{n_i}^{(1)} + \bar{\Delta}_n^{(2)}
$
where in the free field case the ${n_i}^{\rm{th}}$ zero is in the
$n^{\rm{th}}$ degeneracy class. 
In terms of the Dirac and momenta labels, we can write the erstwhile
2--fold degenerate lowest zeroes as
\begin{equation}
 \eta_{\alpha}(\pm p_1,0) = \eta_\alpha^{(0)}(|p_1|,0) 
 \pm \eta_{\alpha}^{(1)}(|p_1|,0) + \bar{\Delta}_\alpha^{(2)}(|p_1|,0)
\quad .
\label{zeroes}
\end{equation}
In this way, the shifts in the positions of the erstwhile two--fold
degenerate zeroes are also determined to ${\cal{O}}(e^2)$
and the ${\cal{O}}(e^2)$--shift is the shift in the average position
of the two zeroes while their relative separation is ${\cal{O}}(e)$.
The calculation of Lee--Yang zeroes is now reduced to the calculation of
the pure gauge expectation values $\Delta_\alpha (p)$ which we do
in Feynman gauge.

%%%%%%%%%%%%%%%%%%%%%%%%%%%%%%%%%%%%%%%%%%%%%%%%%%%%%%%%%%%%%%%%%%%%%%%%
%%%%%%%%%%%%%%%%%%%%%%%%%%%%%%%%%%%%%%%%%%%%%%%%%%%%%%%%%%%%%%%%%%%%%%%%
\paragraph{Results and Conclusions}
%\setcounter{equation}{0}
%%%%%%%%%%%%%%%%%%%%%%%%%%%%%%%%%%%%%%%%%%%%%%%%%%%%%%%%%%%%%%%%%%%%%%%%
%%%%%%%%%%%%%%%%%%%%%%%%%%%%%%%%%%%%%%%%%%%%%%%%%%%%%%%%%%%%%%%%%%%%%%%%

The main plot in Figure 1 is a finite--size scaling
plot for the lowest zero for $\beta=10$ and 
$N=8,10,...,62$ coming from our weak coupling analysis (circles).
The corresponding data from the numerical analysis of \cite{AzDi96}
are also included (crosses). The two  lines are
the linear fit to the lowest zeroes
for lattice sizes $16,20,24$ of \cite{AzDi96}
and a fit to the second zero. These yield slopes $-1.5$ and $-1.4$
respectively, corresponding to $\nu \approx 2/3$.
The insert contains a similar plot at $\beta = 5$ where the circles 
are our
results for $N=2,4,...,42$ and the squares are the results of
\cite{GaLa94} for the lowest zero for
$N=2,4,8$. The line is the best linear fit
to this latter data and yields $\nu$ compatable with $1$.
Fitting to our small lattice  data also
gives $\nu \approx 1$.

Therefore, confining
the finite--size scaling
 analysis to the range $N=2$ -- $8$ yields a slope compatable
with $\nu=1$ in agreement with \cite{GaLa94}. 
A corresponding analysis with
$N=8$--$24$ gives a steeper slope, compatable with $\nu=2/3$
in agreement with \cite{AzDi96}.
It is clear from the figure, however, that
the curve does not in fact settle to a  finite--size scaling line.
Instead as $N$ increases, the lowest zero crosses the
real $1/2\kappa$ axis. The second zero exhibits the same behaviour
as demonstrated by the upper curve in Figure~1. 
The first two zeroes therefore fail to accumulate 
and do not contribute to critical behaviour \cite{LY,Abe,Sa94}.
One interprets these zeroes 
as isolated singularities of thermodynamic functions, 
having measure zero in the thermodynamic limit.

\begin{figure}[htb]
\vspace{13cm}               
\includegraphics{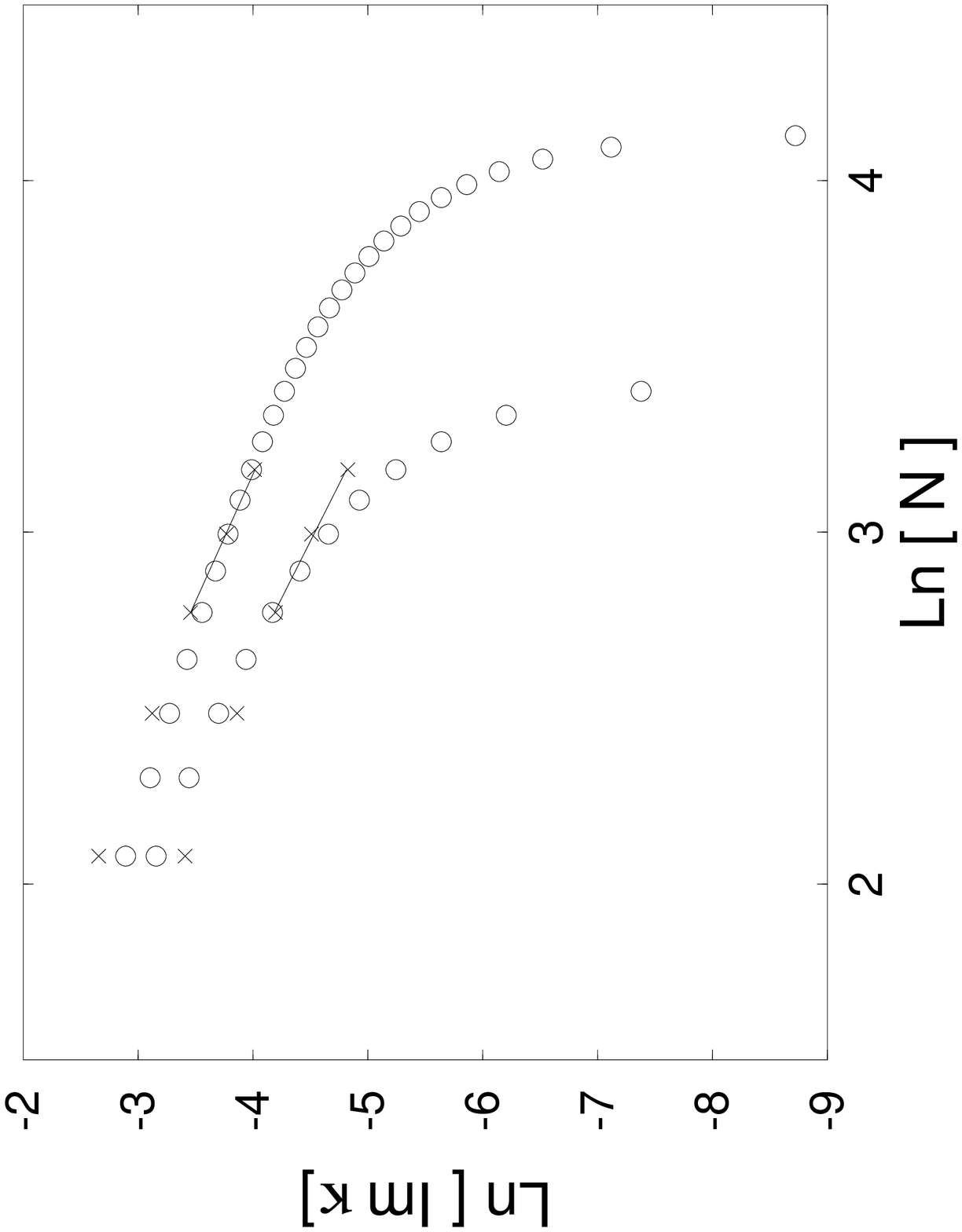}
\includegraphics{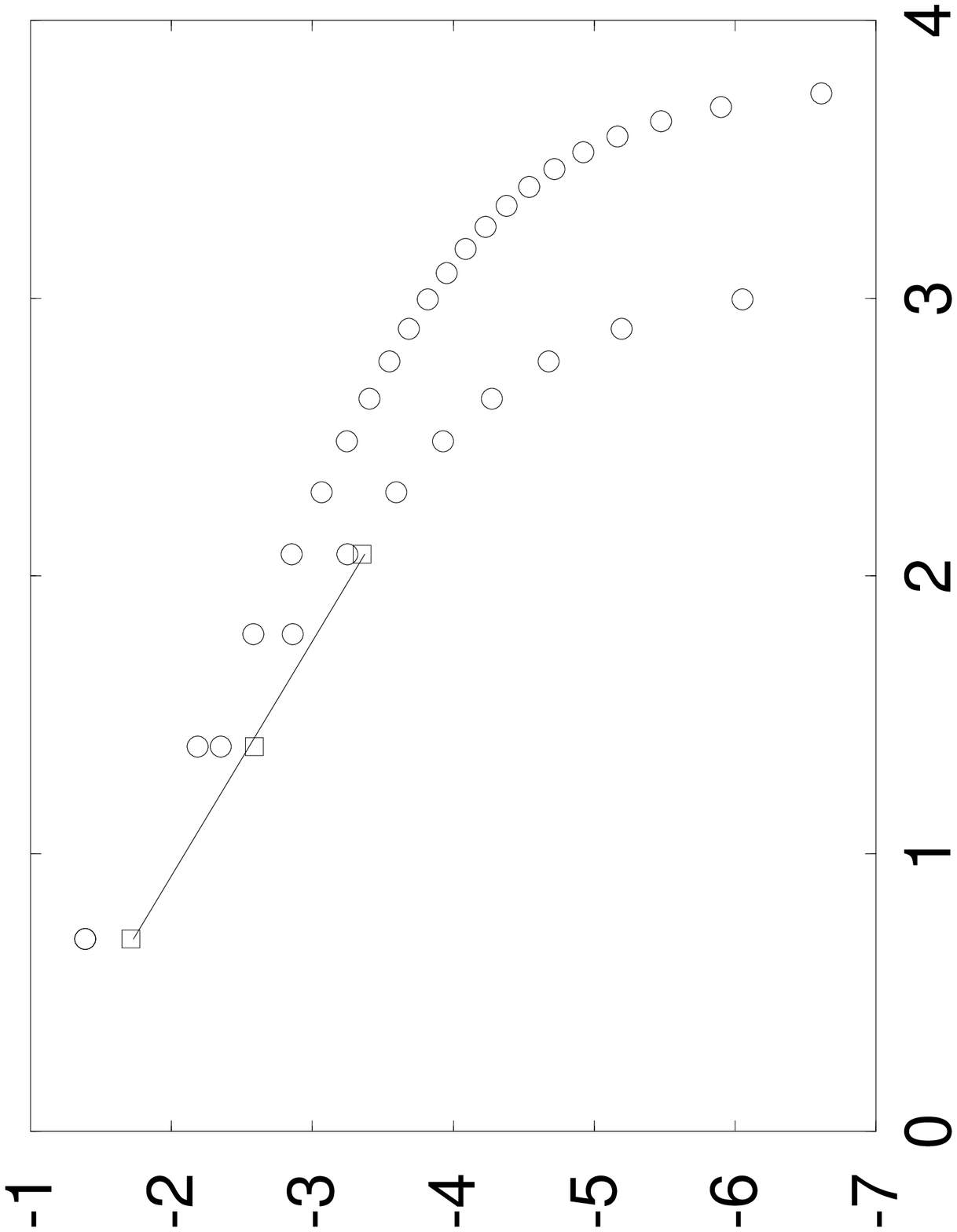}
\caption{The finite--size scaling behaviour of the imaginary
parts of the first two Lee--Yang zeroes for $\beta=10$.
The symbols $\circ$ are our weak coupling expansion results.
The numerical results of \cite{AzDi96} for $N=8, \dots , 24$
are  $\times$  and the lines are 
best fits to these points. The insert is a similar plot for 
$\beta = 5$.
There the \footnotesize $\Box$  \normalsize  are the lowest zeroes
of \cite{GaLa94}.}
\label{fig1}
\end{figure}

It would be extremely difficult
 to detect the isolated nature of these Lee--Yang
zeroes from numerical data alone. A standard numerical technique for
determination of the phase diagram -- assuming accumulation of 
zeroes --
is to approximate the infinite volume critical point by 
the real part of the lowest zero (then a pseudocritical point) for some
reasonably large lattice. Plotted against $\beta$ this
appproximates the phase diagram.
In Figure~2 we present such a plot for $N=24$ (circles)
to compare with the results of \cite{AzDi96} (crosses) also at $N=24$.
The phase diagram of \cite{HiLa98}
for the Schwinger model with {\em{two}} fermion flavours coming from
a separate PCAC based analysis is also included
(squares)
for comparison.

\begin{figure}[htb]
\vspace{13cm}               
\includegraphics{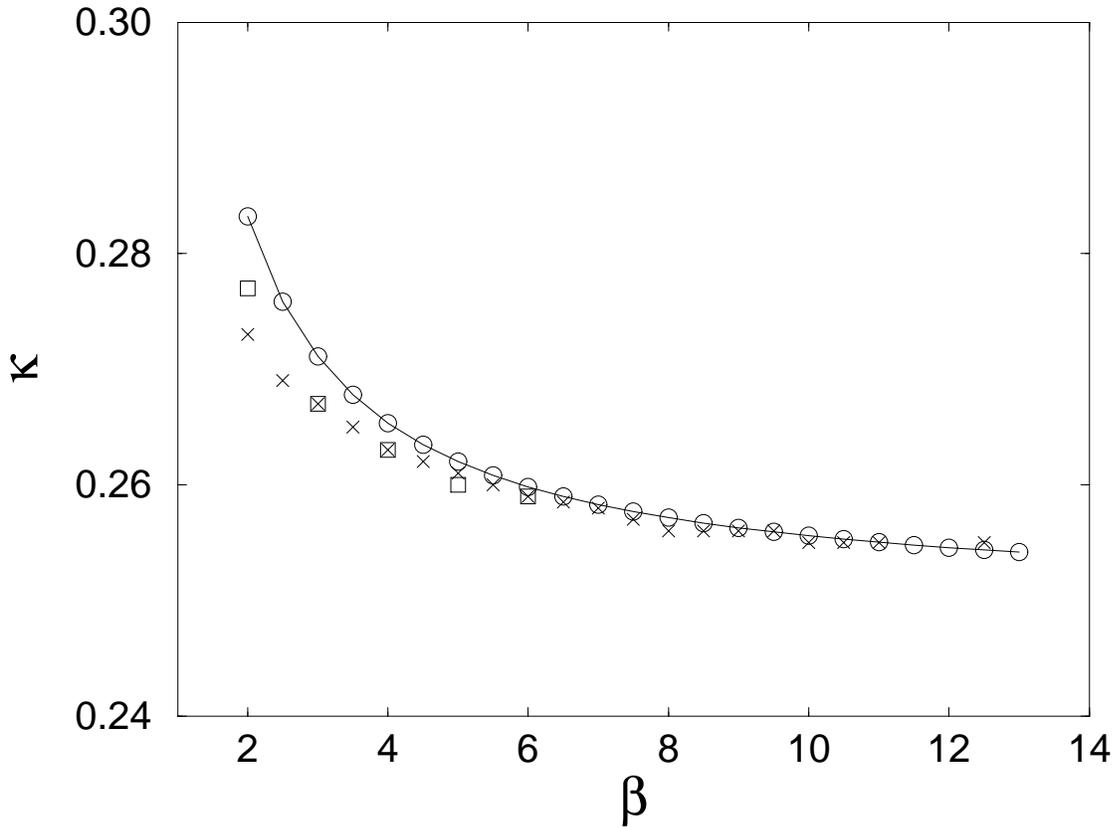}
\caption{The would--be phase diagram coming from the real parts
of the lowest zeroes at various $\beta$ at fixed $N$. 
The symbols $\circ$ are our weak coupling expansion 
results for $N=24$ and  $\times$
are numerical results of \cite{AzDi96} also for $N=24$.
The symbols \footnotesize $\Box$\normalsize  ~ 
are the numerical results of \cite{HiLa98} for $N=16$
and two fermion flavours.}
\label{fig2}
\end{figure}

Now, the ${\cal{O}}(e)$ shift in the free field lowest zero is, 
in fact, 
$
\eta_{\alpha}^{(1)} (\pm 1/2,0) 
 = 
\pm
 ( i / (Na)^d) 
 \sqrt{
  \langle 
     |\phi_\mu(1,0)|
  \rangle
}
=
 \pm i /( 2 \sqrt{\beta} N|\sin{p_1a}|)
$,
the leading finite--size scaling  behaviour of which is 
$N$--independent.
The crossing of the real axis by the first two zeroes is therefore
due to the $N$ dependency of their ${\cal{O}}(e^2)$ average. 
In the four fold degenerate case, the leading $N$ behaviour of 
the zeroes
is again determined by that of their average. 
The four fold degenerate ${\cal{O}}(e^2)$ averages of small momentum
zeroes also cross the 
real axis and do not accumulate.

Although we cannot logically exclude the possibility that some zero
which was far from the real axis in the free fermion case (or some
combination of such zeroes) conspires to accumulate on the real 
$1/2\kappa$ axis in the usual way, such a situation seems unlikely.
Nor can we exclude the possiblilty that just as the subdominant
terms in the weak coupling expansion
 (the ${\cal{O}}(e^2)$ averages) correspond to the dominant
$N$ behaviour, so too could higher order terms in the weak 
coupling expansion
 correspond to
superdominant $N$ effects. This seems again unlikely as the average
zeroes coming from our weak coupling expansion
 agree very well with those of \cite{AzDi96} 
(see \cite{us} for more details). Moreover, superdominant $N$ effects
from subdominant weak coupling terms would spell disaster for lattice
(and indeed continuum) perturbation theory and is an unlikely 
scenario
in this -- a superrenormalisable -- model.

Although the possibility of existence of isolated singularities
and non--accumulation of partition function zeroes has been
known for a long time \cite{LY,Abe,Sa94}, this is to our knowledge 
the first instance 
where such behaviour has been observed. 
In conclusion, in the free case, there is a phase transition
precipitated by the accumulation of Lee--Yang zeroes on the 
real hopping
parameter axis. In the weakly coupled regime at fixed $\beta$, 
this accumulation no longer occurs. 
Instead, the movement of zeroes for  small and
moderately sized lattices mimics phase transition like behaviour.
As the lattice size becomes large, however, 
these zeroes move across the real axis, 
and do not give rise to a phase transition.

{\bf{Acknowledgements}}: We would like to thank the following
for discussions. N. Christ, R. Mawhinney, M.P. Fry, H. Gausterer,
I. Hip, A.C. Irving, C.B. Lang, R. Teppner.

%---------------------------------------------------------

\end{document}